\documentclass[aps,prl,showpacs,twocolumn]{revtex4-2}
\usepackage{graphicx}
\usepackage{subfig}
\usepackage{tabularx}
\usepackage{bm}
\usepackage{amsmath}
\usepackage{esvect}

\providecommand{\abs}[1]{\left\lvert#1\right\rvert}

\providecommand{\bra}[1]{\langle #1 \rvert}
\providecommand{\ket}[1]{\lvert #1 \rangle}

\providecommand{\be}{\begin{equation}}
\providecommand{\ee}{\end{equation}}
\providecommand{\ba}{\begin{eqnarray}}
\providecommand{\ea}{\end{eqnarray}}

\newcommand{\defeq}{\overset{\text{\tiny def}}{=}}
\usepackage{mathbbol}

\usepackage{tikz}
\usepackage{amsfonts,amssymb}
\usepackage{dsfont}
\usepackage{physics}
\usepackage{tocvsec2}

\usepackage{graphicx}
\usepackage{subfig}
\usepackage{tabularx}
\usepackage{bm}
\usepackage{amsmath}
\usepackage{esvect}

\usepackage{siunitx}
\usepackage{hyperref}
\usepackage[capitalize]{cleveref}
\DeclareMathOperator{\sinc}{sinc}

\providecommand{\abs}[1]{\left\lvert#1\right\rvert}

\providecommand{\bra}[1]{\langle #1 \rvert}
\providecommand{\ket}[1]{\lvert #1 \rangle}

\providecommand{\be}{\begin{equation}}
\providecommand{\ee}{\end{equation}}
\providecommand{\ba}{\begin{eqnarray}}
\providecommand{\ea}{\end{eqnarray}}
\usepackage{amsfonts,amssymb}
\usepackage{dsfont}
\usepackage{physics}
\usepackage{tocvsec2}

\hypersetup{colorlinks=true,linkcolor=blue,citecolor = blue, urlcolor=black}
\newcommand{\beq}{\begin{equation}}
\newcommand{\eeq}{\end{equation}}

\usepackage{appendix}

\begin{document}

\title{Approaching maximal precision of Hong-Ou-Mandel interferometry with non-perfect visibility}
 
\author{O. Meskine$^{1}$}
\thanks{These two authors contributed equally}
\author{ E. Descamps$^{1,2}$}
\thanks{These two authors contributed equally}
\author{ A. Keller$^{1,3}$}
\author{A. Lemaître$^4$}
\author{ F. Baboux$^1$}
\author{ S. Ducci$^1$}
\email{corresponding author, experiments: sara.ducci@u-paris.fr}
\author{P. Milman$^{1}$ }
\email{corresponding author, theory: perola.milman@u-paris.fr}

\affiliation{$^{1}$Laboratoire Mat\'eriaux et Ph\'enom\`enes Quantiques, Universit\'e Paris Cit\'e, CNRS UMR 7162, 75013, Paris, France}
\affiliation{$^{2}$Département de Physique de l’Ecole Normale Supérieure - PSL, 45 rue d’Ulm, 75230, Paris Cedex 05, France}
\affiliation{$^{3}$Department de Physique, Université Paris-Saclay, 91405 Orsay Cedex, France}
\affiliation{$^{4}$Univ. Paris-Saclay, CNRS, Centre de Nanosciences et de Nanotechnologies, 91120 Palaiseau, France}

\begin{abstract}
In quantum mechanics, the precision achieved in parameter estimation using a quantum state as a probe is determined by the measurement strategy employed.  The  quantum limit of precision is bounded by a value set by the state and its dynamics. Theoretical results have revealed that in  interference measurements with two possible outcomes, this limit can be reached under ideal conditions of perfect visibility and zero losses. However, in practice, these conditions cannot be achieved, so precision {\it never} reaches the quantum limit. But how do experimental setups approach precision limits under realistic circumstances? In this work we provide a model for precision limits in two-photon Hong-Ou-Mandel interferometry using coincidence statistics for non-perfect visibility and temporally unresolved measurements. We show that the scaling of precision with visibility depends on the effective area in time-frequency phase space occupied by the state used as a probe, and we find that an optimal scaling exists. We demonstrate our results experimentally for different states in a set-up where the visibility can be controlled and reaches up to $99.5\%$. In the optimal scenario, a ratio of $0.97$ is observed between the experimental precision and the quantum limit, establishing a new benchmark in the field.

\end{abstract}
\vskip2pc 
 
\maketitle

The Hong-Ou-Mandel (HOM) interferometer is currently used to demonstrate the phenomenon of bunching of two identical, independent bosonic quantum particles, such as single photons \cite{PhysRevLett.59.2044} (see \cref{fig1}). In this setup, photons are made to interfere on a balanced beam-splitter (BS) and their detection in coincidence at the output indicates whether they have bunched or not. To control the distinguishability of the two paths of the interferometer, a time delay can be introduced for one of the input photons, consequently changing the coincidence detection probability. Despite its seemingly straightforward operating principles, the HOM interferometer has found diverse applications beyond its original scope \cite{EPJDUs}: the coincidence detection signal at the BS output has been demonstrated to serve as an entanglement witness \cite{brecht_characterizing_2013, eckstein_broadband_2008, vanEnk}, to provide phase space information about the spectral function \cite{douce_direct_2013, PhysRevLett.115.193602, Kurzyna:22}, and to enable the simulation of different quantum exchange statistics \cite{Anyons, francesconi_engineering_2020}, among other applications \cite{HOM1, HOM2, PhysRevA.85.043816}. In particular, the HOM interferometer is a valuable apparatus for quantum parameter estimation both in the time unresolved measurement (TUM) regime \cite{PhysRevA.108.013707, fabre_parameter_2021, Lundeen, lyons_attosecond-resolution_2018, chen_hong-ou-mandel_2019, PhysRevApplied.19.054092, Kwiat} and in the time resolved measurements (TRM) one \cite{Scott}: its low-intensity regime opens the possibility of applying the tools of quantum metrology to small and fragile probes, as biological ones \cite{Bio}; since the HOM effect is based on two-photon interference, it is robust against background noise, group velocity dispersion \cite{KwiatVelho} and phase perturbations \cite{NoiseHOM}. Last but not least, theoretical and experimental results indicate that it can arbitrarily approach the quantum precision limit for time delay (or path difference) estimation \cite{chen_hong-ou-mandel_2019, Kwiat,fabre_parameter_2021,PhysRevA.108.013707, Scott}. However, in spite of the recent experiments reaching up to attosecond precision on time delay estimations \cite{lyons_attosecond-resolution_2018, chen_hong-ou-mandel_2019,Kwiat,Lundeen, fringes}, the mechanisms determining the limits and limitations of different quantum states for time measurement precision using the HOM with respect to the maximal achievable precision are unknown under realistic conditions of non perfect visibility. In particular, such limitations depend on the adopted measurement strategy. In the present manuscript we will concentrate on the TUM regime with unresolved photon number detection where, if losses are ignored, photons either bunch or anti-bunch at each run of the experiment, leading to a two-possible outcome experiment \cite{SM}. A curious result observed not only for HOM interferometers but also for any parameter estimation protocol based on dichotomic measurements (see also \cite{LuizExp, Dalvit_2006}, for instance) concerns the behavior of precision with the visibility $V$ at the point where, in the ideal case, the former is expected to saturate the quantum limit. For instance in the HOM set-up, if $V=1$, one can attain the quantum precision limit at zero delay, where photons either perfectly bunch or anti-bunch. Nevertheless, in the experimentally realistic case where the visibility $V <1$, total bunching or anti-bunching is no longer observed and  precision drops down to zero at this time delay. A way to circumvent this in the context of a Mach-Zender interferometer was studied in \cite{Banaszek} using a mode engineering-based strategy. As for the HOM experiment, for finite visibility, the quantum precision limit can only be approached, and the maximal attainable precision will occur for a finite delay between photons, as discussed later in this paper. However, it is not clear if and how the maximal achievable precision depends on the probe's state spectral function in a HOM experiment for a given visibility \cite{fabre_parameter_2021,lyons_attosecond-resolution_2018, chen_hong-ou-mandel_2019, PhysRevApplied.19.054092}.

\begin{figure}[h]
    \includegraphics[width=\columnwidth]{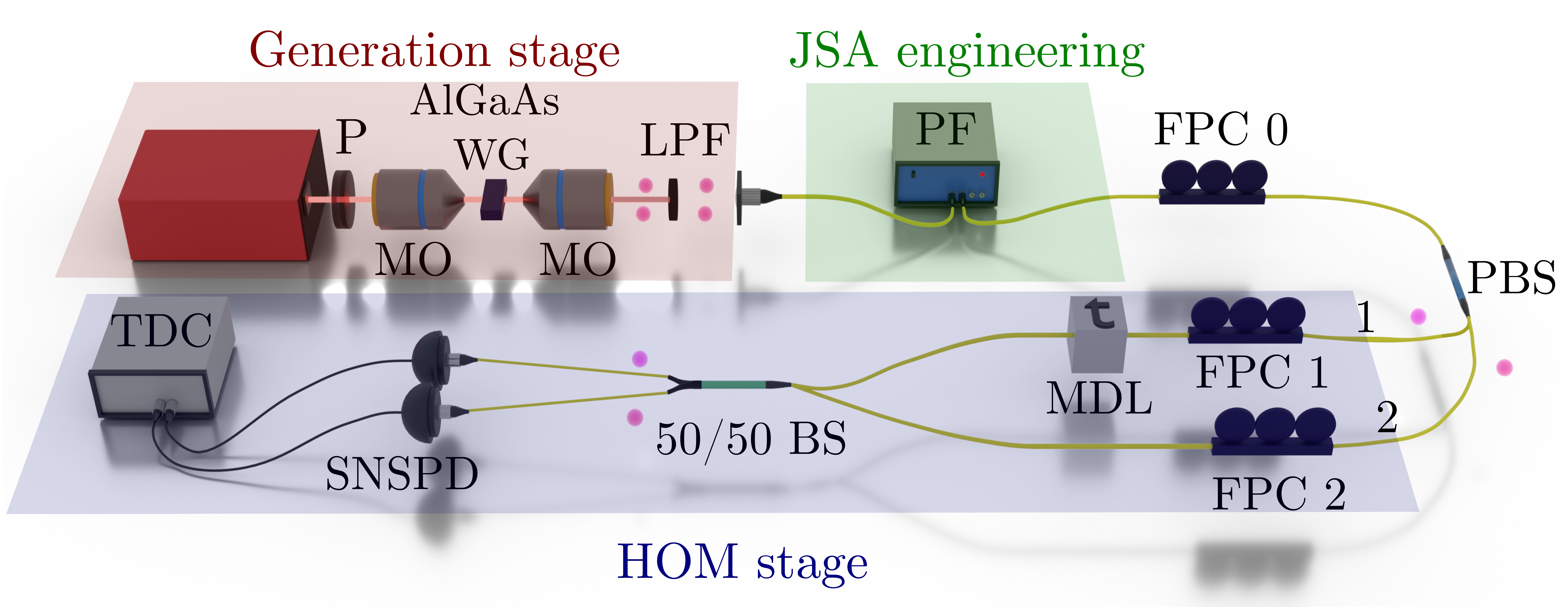}
    \caption{Experimental setup for investigating the metrological performance of the Hong  Ou Mandel (HOM) experiment, showing the generation, joint spectral amplitude (JSA) engineering and HOM interferometer stages}
    \label{fig1}
\end{figure}
In the present Letter, we provide a theoretical model and its experimental demonstration explaining both qualitatively and quantitatively the performance of the HOM experiment as a quantum metrological tool, as a function of the input state and visibility. For such, we have exactly theoretically predicted and experimentally confirmed the wavefunction-dependency of the scaling of the maximal precision with the visibility for different wavefunctions and interpreted our results in terms of phase space properties. For some configurations, we reach the highest ratio between the achieved precision and the maximum possible one to date, thereby setting a new benchmark in this field. 

In a typical metrological protocol, a probe is prepared in an initial state, and undergoes a dynamical evolution depending on a parameter to be estimated, $\theta$. The probe is measured providing an outcome $k$, which is used to estimate $\theta$. By associating the function $p_k(\theta)$ to the probability of obtaining an outcome $k$, the precision on the estimation of $\theta$ is bounded by the relation $\delta \theta \geq 1/\sqrt{\nu F(\theta)}$, where $F(\theta)= \sum_k \frac{1}{p_k(\theta)}\left ( \frac{\partial p_k(\theta)}{\partial \theta} \right )^2$ is the Fisher information (FI) and $\nu$ is the number of repetitions of the experiment. When using quantum mechanical resources - as individual photons, which is the case in a HOM experiment - one can define the quantum Fisher information (QFI) by using a quantum state as a probe \cite{PhysRevLett.96.010401}. The probe's evolution depends on the parameter $\theta$, and for pure states and unitary evolutions, this dependency can be expressed as $\ket{\psi(\theta)}=e^{i\hat H \theta}\ket{\psi}$, where $\hat H$ is the Hamiltonian generating the dynamical evolution. Precision is thus limited by the relation $\delta \theta \geq 1/\sqrt{\nu {\cal F}}$, where ${\cal F}$ is the QFI, obtained by maximizing $F(\theta)$ over all possible measurements on $\ket{\psi(\theta)}$. In the case discussed above, ${\cal F}= 4 \Delta^2 \hat H$, where the variance is taken with respect to the initial state $\ket{\psi}$ \cite{VarianceQFI}. The bound $1/\sqrt{\nu {\cal F}}$ is called the {\it quantum Cramér-Rao bound} (QCR) \cite{Cramer}, the quantum precision limit \cite{True}. 

 There is no general rule for finding an experimental measurement strategy where $F(\theta) ={\cal F}$, even though optimization procedures can be applied to particular states \cite{pinel_ultimate_2012, gessner2022quantum} and symmetry arguments can be evoked in specific situations \cite{PhysRevA.79.033822, PhysRevA.87.043833}, as in the HOM experiment \cite{Lundeen, PhysRevA.108.013707}. As shown in the literature, time precision in HOM interferometry with perfect visibility can reach the QCR bound \cite{chen_hong-ou-mandel_2019,Kwiat, fabre_parameter_2021, Lundeen}. Consequently, several attempts have been made to reach this bound, obtaining astonishing precision on the estimation of time delays in HOM experiments  \cite{lyons_attosecond-resolution_2018, chen_hong-ou-mandel_2019, PhysRevApplied.19.054092, Kwiat}.  
To analyze these results, we consider as initial state (probe) a photon pair prepared in an arbitrary pure state, that enters the two input arms $1,2$ of a perfectly balanced BS:
\be\label{state}
\ket{\psi}=\int \int d\omega_1 \omega_2 f(\omega_1, \omega_2)\hat a_1^{\dagger}(\omega_1)\hat a_2^{\dagger}(\omega_2)\ket{0},
\ee
where $f(\omega_1,\omega_2)$ is the complex-valued normalized joint spectral amplitude (JSA). Before impinging the balanced BS, one of the photons in \eqref{state} is subjected to a time delay $\tau$, the parameter to be estimated. This delay is described by a unitary evolution associated to the Hamiltonian $\hat H=\hbar \int d\omega  \omega \hat a_1^{\dagger}(\omega) \hat a_1(\omega)=\hbar \hat \omega_1$ (we have supposed that arm $1$ is delayed). State \eqref{state} becomes $\ket{\psi(\tau)}= e^{i\hat H \tau/\hbar}\ket{\psi}=\hat U\ket{\psi}$. After the BS, the probability of detecting both photons in coincidence $P_c (\tau)=\frac{1}{2}(1-\bra{\psi}\hat U^{\dagger} \hat S \hat U \ket{\psi})$ \cite{PhysRevA.108.013707}, where $\hat S \hat a_1^{\dagger}(\omega_1)\hat a_2^{\dagger}(\omega_2)\hat S^{\dagger}=  \hat a_1^{\dagger}(\omega_2)\hat a_2^{\dagger}(\omega_1)$ is a swap of spatial modes. $P_c(\tau)$ is typically directly obtained from the recorded experimental data. $P_a(\tau)$, the probability of anti-coincidences, can either be inferred from direct detection \cite{Kwiat} or using $P_a(\tau)=1-P_c(\tau)$ \cite{Note3}. Finally, the QFI can be expressed as \cite{SM} ${\cal F}=4 \Delta^2\hat \omega_1$. For perfectly symmetric (S) (anti-symmetric (AS)) states $\ket{\psi}_{S(AS)}$ with respect to the exchange of spatial modes, $\hat S \ket{\psi}_{S(AS)}= \pm  \ket{\psi}_{S(AS)}$ and $P_c(\tau=0) = 0(1)$. Hence, the HOM has perfect visibility and the QFI can be reached.

The HOM experiment provides information about the collective variables $\omega_-= \omega_1 -\omega_2$ \cite{Lundeen}. We will consider that the input photons of the interferometer are generated by Spontaneous Parametric Down Conversion (SPDC) and that  $f(\omega_1, \omega_2)=f_-(\omega_-)f_+(\omega_+)$ in \eqref{state}, with $\omega_{\pm}=\omega_1\pm\omega_2$. Functions $f_-$ and $f_+$ are normalized functions related to the phase matching condition and to the energy conservation, respectively. 
The best configuration for metrology is the one where $f_+$ is a Dirac function centered on $\omega_p$ (the pump's frequency) and $\omega_+$ is close to constant (strict energy conservation), maximizing the frequency correlation between photons \cite{MacconeNature, PhysRevLett.131.030801}. Therefore, supposing that both photons have the same spectral variance, ${\cal F} = 4\Delta^2 \omega_1 = \Delta^2 \omega_-$. In addition, $\bra{\psi}\hat U^{\dagger} \hat S \hat U \ket{\psi} = \int  d\omega_- e^{i\omega_-\tau}f_-(\omega_-)f_-^*(-\omega_-) = W(0,\tau)$. Here, $W(\mu,\tau)$ denotes the chronocyclic Wigner function associated with $f_-(\omega_-)$ on the time-frequency phase space (TFPS), specifically on the axis $\mu=0$ while $\tau$, the time delay, is variable. Hence, $\mu$ is the phase space variable associated with $\omega_-$. It was shown in \cite{douce_direct_2013} and experimentally validated in \cite{PhysRevLett.115.193602, Kurzyna:22} that the HOM experiment directly measures the Wigner function points $W(0,\tau)$, {\it i.e.}, along the axis $\mu=0$ of TFPS. Adopting this representation facilitates an intuitive understanding of the factors that dictate the limitations imposed by non-perfect visibility, which we study in the following.


To this aim, we model the dependency of  $P_c(\tau)$ with the visibility $V$ as \cite{SM}:
\be \label{lambda}
P_c(\tau)=\frac{1}{2}-\frac{V}{2}W(0,\tau),
\ee
where $0 \leq V \leq 1$ and $W(0,\tau)$ is the Wigner function of a perfectly symmetric state  \cite{CommentOdd}, so $W(0,0)=1$ and $P_c(0)=(1-V)/2$, defining the visibility. We are not considering explicitly the role of experimental noise or losses in the measurement, evolution or preparation steps: they can either be included in the QFI - which is consequently modified - or in the state's purity \cite{LuizMetro, Outro}. Including noise in the state corresponds to considering a different (non-pure) state as a probe, so a different function $W(0,\tau)$. Thus, state noise or measurement losses have no incidence on the model \eqref{lambda}, that remains valid. In addition, loss and visibility are independent quantities and even for pure states in a lossless configuration a non-unit visibility can be observed due to state preparation imperfections that cannot be circumvented \cite{SM}. Hence, we will focus exclusively on the visibility in the present work. Using \eqref{lambda}, the FI at point $\tau$ is given by 
\be\label{fishmax}
F(V, \tau)=V^2\frac{(W'(0,\tau))^2}{(1-V^2 W^2(0,\tau))},
\ee
where $'$ denotes the time derivative. By defining ${\rm max}_{\tau} F(V, \tau) \defeq \widetilde F_{V}$, we obtain, for $V=1$ (perfectly S or AS states), $\widetilde F_{1} = F (1,0)={\cal F}$  \cite{Lundeen, PhysRevA.108.013707,  PhysRevA.87.043833}. This shows that the quantum precision limit can be achieved for perfect visibility. For $V \neq 1$ we can still compute $\widetilde F_{V}$, which is obtained at a point $\tau = \tau_M \neq 0$. As a matter of fact, for $\tau=0$,  the function  $F(V, 0)$, when considered as a function of $V$, exhibits a discontinuity at $V=1$: indeed, $F(V <1,0) = 0$, while $F(1,0) = \widetilde F_{1} =  \mathcal{F}$, as previously shown. For $V <1$, the maximal values of $\widetilde F_{V<1}$ satisfying $\widetilde F_{V<1}= -W''(0,\tau_M)/W(0,\tau_M)$. Experimental investigations of these results were conducted in \cite{Kwiat, lyons_attosecond-resolution_2018, PhysRevApplied.19.054092}, and \cite{LuizExp}, but the overall behavior of the attainable values of $\widetilde F_{V}$ and how their limitations and their scaling with visibility is related to the state's wave-function remains unknown. We will now elucidate how $\widetilde F_{V}$ approaches ${\cal F}$. Importantly, we find that this approach depends not only on the visibility but we also identify a relation with the effective phase space occupation of the state's Wigner function $W(\mu, \tau)$. Indeed, the scaling of $\widetilde F_{V}$ with $V$ is connected to how far the quantum state is from saturating the time-frequency Heisenberg uncertainty principle \cite{PhysRevA.104.L050204, PhysRevA.105.052429}.  A first remark is that since $V \leq 1$, Eq.  \eqref{fishmax} leads to $F(V,\tau) \leq V^2  F(1,\tau)$, so $\widetilde F_{V}=V^2 {\cal F}$ is the best possible scaling of the FI with $V$. This is a proof that for $V < 1$ the HOM can never reach the QCR bound, even in the absence of losses. In addition, using \eqref{fishmax} we see that the best possible scaling is obtained when $W(0,\tau_M)=0$ ({\it i.e.}, $P_c(\tau_{M})=1/2$) and $W'(0,\tau_M)\neq 0$. A sinusoidal function of frequency $\sqrt{{\cal F}}$ satisfies these conditions (a solution also leading to a constant FI in $\tau$ for $V=1$  \cite{PhysRevE.97.042110, SM}). This unphysical solution represents the limit situation of states occupying a large effective area in TFPS  \cite{Zurek}, as Schrödinger cat (SC)-like states \cite{PhysRevLett.103.253601, Kwiat, chen_hong-ou-mandel_2019} with $\Delta^2\hat \omega_- \Delta^2 \hat t \gg 1$ \cite{SM}. Surprisingly, SC-like states exhibit remarkable robustness in the presence of decreased visibility, making them the most resilient states in HOM-based quantum metrology, which is yet another interesting quality of these states in quantum metrology \cite{Comment2}.

States leading to the worst possible scaling of $\widetilde F_{V}$ with $V$ minimize $\widetilde F_{V<1} ~ \forall ~V$. The quantity $-W''(0,\tau)$ in the region $\tau \ll \Delta \hat \omega_-$ is minimized by Gaussian states \cite{PhysRevA.72.052332, Toppel_2012}. Consequently, in this region (where lies the value of the parameter to be estimated), Gaussian states also minimize $W'(0,\tau)$, so $W(0,\tau)$ is maximal. For this reason, they are the states exhibiting the worst scaling of $\widetilde F_{V<1}$ with $V$, even when they have the same limit value for the FI as SC-like states for $V=1$ and $\tau=0$, {\it i.e.}, the QFI. Interestingly, their scaling with $V$ does not depend on the values of $\Delta \hat \omega_-$ or $\Delta \hat t$, but on the associated function (Gaussian), which is univocally determined by the product of the two quantities. As states' phase space occupation change from the Gaussian to the sinusoidal behavior, their scaling with visibility improves. Interestingly, while the wavefunction shape {\it do not} play a role in the value of ${\cal F}$ \cite{PhysRevLett.131.030801, SM} it does play a role in the scaling of $\widetilde F_{V}$ with $V$, in a way that is related to the effective occupation of the TFPS. Previous works \cite{Zurek} established the connection between metrological properties and the state's occupation of the quadrature phase space \cite{PhysRevLett.122.040503, PhysRevX.8.041038}, where small structures determine quantum properties such as the QFI. Our analysis indicate that such structures also contribute to the optimization of the scaling of precision with visibility. We emphasize that our analysis is applicable to various experimental setups where the model \eqref{lambda} holds \cite{LuizExp, Dalvit_2006}, such as experiments with more than two photons \cite{doi:10.1142/S0217979207038186}, where the scaling with $V$ is important to determine the tolerance of the sub-shot noise region to visibility decrease (see \cite{LuizExp, SM}).

We now validate our model using an experiment allowing to engineer two-photon states described by different functions $f_-(\omega_-)$ exhibiting diverse scaling behaviors with the visibility $V$. 
The quantum source consists of an AlGaAs Bragg reflector waveguide, generating polarization-entangled photon pairs via type II Spontaneous Parametric Down-Conversion (SPDC) at telecom wavelengths and operating at room temperatures \cite{Appas2022}. A sketch of the experimental setup is provided in Fig. \ref{fig1}. A continuous-wave laser having a wavelength $\lambda_{pump} = \SI{772.42}{\nano \meter}$ is coupled into the waveguide using a microscope objective (MO). The output signal is collected by a second microscope objective, and the pump beam is filtered out using a long-pass filter (LPF). The generated photon pairs are then collected in a single mode fiber and possibly directed to a programmable filter (PF, Finisar 4000s), enabling the JSA engineering. When the filter is not inserted, the state generated by the source is described by $f_-(\omega_-)=\sinc(a\omega_-^2+b\omega_-+c)$
where the coefficients $a$, $b$ and $c$ are related to optical properties of the material such as birefringence and chromatic dispersion \cite{malte2019,SM}.

In addition to the study of this case, three different filter shapes are used: a $\SI{15}{\nano\meter}$-wide rectangular filter centered on the degeneracy wavelength $\lambda_{deg}=\SI{1544.8}{\nano\meter}$, a Gaussian filter of identical width centered at the same wavelength, and a combination of two $\SI{5}{\nano\meter}$-wide rectangular filters centered at $\lambda_1=\SI{1560}{\nano \meter}$ and $\lambda_2=\SI{1530}{\nano \meter}$ corresponding to energy-matched channels and allowing to create a SC-like state, analogously to as described in \cite{OuMandel, fabre_producing_2020}. The functions $f_-$ and $W(0,\tau)$ associated with the four states are presented in Table.1. They can be classified according to the parameter ${\cal S} = \Delta^2 \hat \omega_- \Delta^2 \hat t$, which determines the scaling with respect to $V$ (see \cite{SM} for details).

 At the output of the filtering process, the photon pairs are separated by a polarizing beam splitter (PBS). The $H$ ($V$)-polarized photon enters the HOM interferometer through the arm 1 (2). Precise control over the polarization distinguishability, and thus the HOM visibility, is enabled by two fibered polarization controllers, one in each arm (FPC1 and FPC2).
 The temporal delay between the two photons is controlled by a motorized optical delay line (MDL). The two paths are recombined and separated by a 50/50 BS, then directed to superconducting nanowires single photon detectors (SNSPD). Temporal correlations between the detected photons are analyzed by a time-to-digital converter (TDC). \\
 We perform a series of measurements on the four states, systematically varying $V$ to investigate the scaling of the ratio $\widetilde F_V/{\cal F}$. \cref{fig2} illustrates the results obtained. The coincidence counts data (red points) are fitted (red lines) using $P_c(\tau)$ of Eq. \eqref{lambda} and the theoretical expression of each wavefunction \cite{SM}. The FI $F(V,\tau)$ (blue lines) is then computed using Eq. \eqref{fishmax}. The error bars associated to the experimental points are estimated assuming Poissonian statistics. 
 \renewcommand{\arraystretch}{2}
\begin{table}[th]
    \begin{center}
        \begin{tabular}{|c|c|c|}
            \hline
            State & $f_-(\omega_-)$ &$W(0,\tau)$ \\ \hline
            Sinc & $\sinc(a\omega_-^2+b\omega_-+c)$ &$\int d\omega_-f_-(\omega_-)f_-^*(-\omega_-)e^{-i\omega_- \tau}$ \\ \hline
            Gauss &$\exp(-\omega_-^2/2\sigma_{\omega_-}^2)$ &$\exp(-\tau^2\sigma_{\omega_-}^2/2)$\\ \hline
            Rect & $\Pi(\frac{\omega_-}{\Delta\omega_-})$& $\sinc(\frac{\Delta\omega_-\tau}{2})$\\ \hline
            SC & $\Pi(\frac{\omega_--\omega'}{\Delta\omega'})+\Pi(\frac{\omega_-+\omega'}{\Delta\omega'})$ & $\sinc(\frac{\Delta\omega'\tau}{2})\cos{(\omega'\tau)}$\\\hline
        \end{tabular}
    \end{center}
    \caption{$f_-$ and $W(0,\tau)$ functions corresponding to the four  states studied. The different variables introduced are defined in the left column of Fig.2}
	\label{tab1}
\end{table}
\renewcommand{\arraystretch}{1}
 \begin{figure}[h]
    \includegraphics[width=\columnwidth]{Figure2.pdf}
    \caption{Left column: Joint Spectral Amplitude of the four different states analyzed in this work. Central and right columns: corresponding Hong-Ou-Mandel coincidence probability $P_c(t)$ and FI for different values of visibility $V$.}
    \label{fig2}
\end{figure}

\begin{figure}[h]
    \includegraphics[width=\columnwidth]{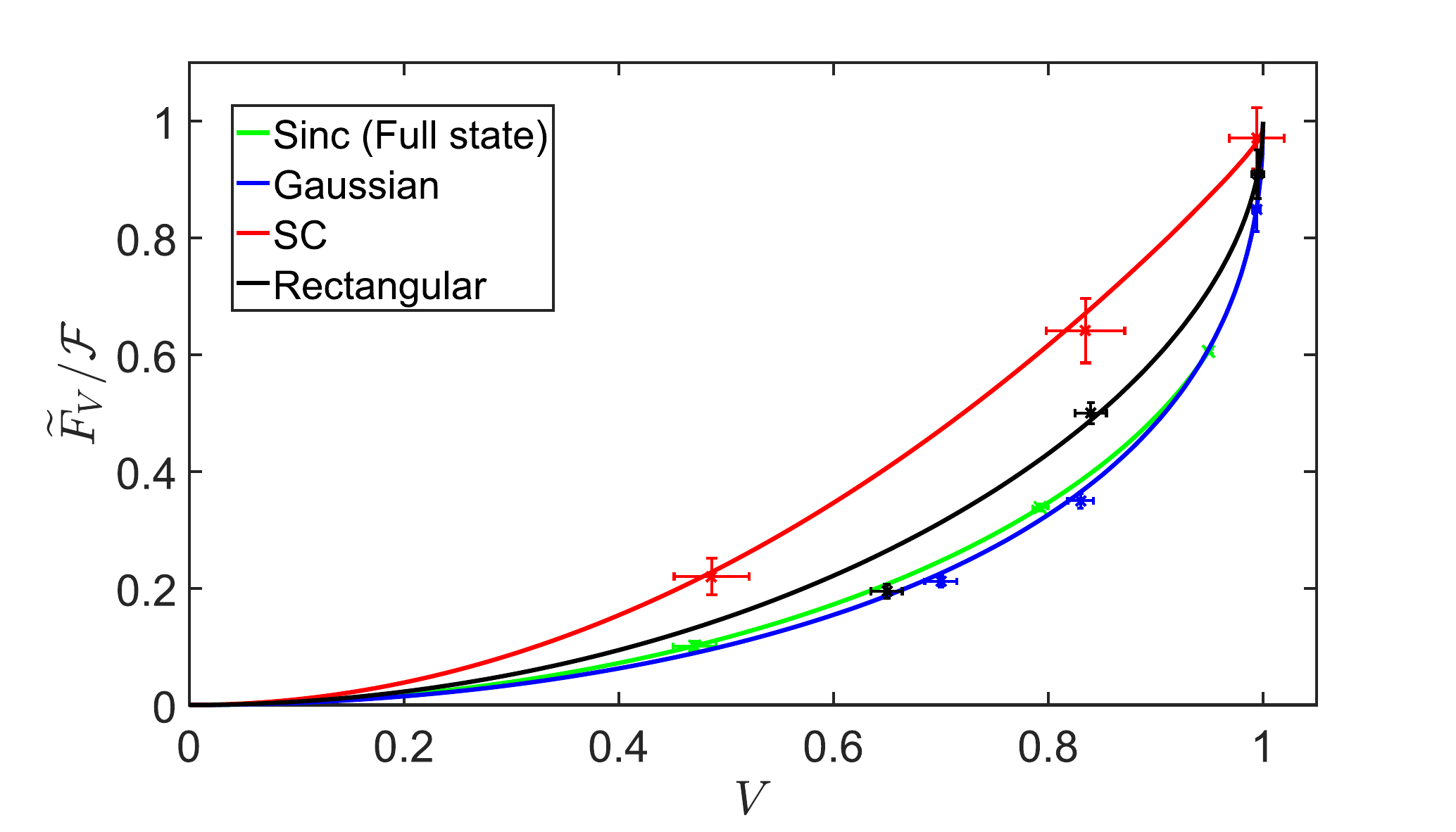}
    \caption{Scaling of the ratio $\widetilde F_{V}/{\cal F}$ for different biphoton states with respect to the HOM visibility $V$. For SC-like states a maximal value of $\widetilde F_{V}/{\cal F}=0.97$ is attained for $V=99.4\%$. Notice that the experimental values of ${\cal F}$ are different for each state.}
    \label{fig3}
\end{figure}

Firstly, we notice that a reduction in visibility leads to a decrease in $F(V,\tau)$. As expected, with finite visibility, the value of $F(V,\tau)$ drops to zero at $\tau=0$. Remarkably high visibilities exceeding $99\%$ are achieved with the Gaussian, rectangular and SC-like states. Due to a small modal birefringence of the AlGaAs source, the maximum visibility attainable with the full state is $94,9\%$, still an excellent value given the broad spectral width it covers, $\approx \SI{100}{\nano\meter}$. This broad spectrum results in a narrow HOM curve $P_c(t)$, leading to a high FI value of $\SI{2100}{\pico\second^{-2}}$, which is two orders of magnitude higher than those obtained with the Gaussian and rectangular states.
 Filtering the quantum state decreases the number of detected photons, therefore influencing the overall performance of the metrological protocol in a given integration time. Nevertheless, in this proof-of-principle experiment, we are mainly interested in testing the scaling of the ratio $\widetilde F_V/{\cal F}=\max_\tau F(V,\tau)/\Delta^2\omega_-$ to demonstrate our model, which can serve as a guideline to other experiments using different strategies. In \cref{fig3}, the evolution of this ratio with respect to visibility for the four engineered states is reported, both for experiments (points) and theory (lines). The error bars associated to the experimental points are estimated by extracting the visibility from the HOM interferogram fit reported in \cref{fig2}. The theoretical plots are obtained by using the functions $f_-$ and $W(0,\tau)$ defined in Table.1. We clearly observe that the SC-like state exhibits the most favorable scaling behavior, in contrast to the Gaussian state, which displays a less optimal one. For instance, at a visibility level of around $99.4 \%$$(83\%)$, the ratio drops to $0.97$ $(0.64)$ for the SC-like state and to $0.85$ $(0.35)$ for the Gaussian state. 

In conclusion, we have presented a well-grounded theoretical model and experimentally confirmed it, showing how precision limits scale with both visibility and the state's wavefunction in practical metrological protocols employing the HOM effect in the broadly used regime of TUM. The very good agreement between the experimental results and the simulations supports the validity of the model given by Eq. \eqref{lambda}. Our findings show that reaching the precision limits in realistic conditions presents challenges that depend on the particular state under consideration. Moreover, our theoretical and experimental analysis establishes a general framework for interpreting previous experimental results \cite{lyons_attosecond-resolution_2018, chen_hong-ou-mandel_2019, PhysRevApplied.19.054092, Kwiat, SampleCats}. We focused on the TUM regime rather in the TRM one, or other measurement strategies based on modal  filtering or post-selection \cite{OuMandel, donohue_quantum-limited_2018, TRNR, ChenChen, SampleCats} because the latter do not depend on the initial wavefunction \cite{fabre_producing_2020}. However, the obtained filtered modes will obey the same behavior as the one detailed in the present work, so these settings can also benefit from our results. 
Our work holds significant implications, particularly in aiding the identification of optimal conditions to advance HOM-based quantum metrology protocols, leading to enhanced precision in measurements while minimizing the number of repetitions.
Finally, some aspects of the presented results can be readily extended to other experiments where parity measurements are employed for quantum parameter estimation \cite{PhysRevLett.78.2547, PhysRevLett.89.200402, douce_direct_2013, PhysRevLett.110.100404, Winkelmann_2022}. 

\section*{Acknowledgements}

We acknowledge funding from the Plan France 2030 through the project ANR-22-PETQ-0006, N. Fabre, G. Bié Alves and J. Lorgeré for fruitful discussions and M. Karr-Ducci for Fig. $1$. O.M. acknowledges Labex SEAM (Science and Engineering for Advanced Materials and devices), ANR-10-LABX-0096 and ANR-18-IDEX-0001 for financial support. 

\onecolumngrid

\section{The role of time resolution: the general case}

We'll set a general frame for coincidence detection in a HOM experiment ranging from the time unresolved case (as ours, and the one of the original HOM experiment), and the time resolved one. For such, we'll express the source's output state in the time basis as $\ket{\psi}= \int \int d t_1 d t_2 JTA(t_1, t_2) \ket{t_1, t_2}$, where $JTA$ is the Joint Temporal Amplitude. The different steps of the HOM interferometer in this basis can be expressed as follows: the time delay in, say, arm $1$, transforms the state as $e^{i\hat \omega_1 \tau} \ket{\psi}= \int \int d t_1 d t_2 JTA(t_1, t_2) \ket{t_1+\tau, t_2}$. So, after the beam splitter, we have that the terms leading to coincidence are, up to a normalization constant, 
\begin{eqnarray}\label{BS}
&&\ket{\psi'}= \int \int d t_1 d t_2 JTA(t_1, t_2)( \ket{t_1+\tau, t_2}+ \ket{t_2,t_1+\tau})=\\
&& \int \int d t_1 d t_2 (JTA(t_1-\tau, t_2)-JTA(t_2,t_1-\tau)) \ket{t_1, t_2} \nonumber
\end{eqnarray}

We'll now model detection in arms $A$ and $B$ (output of the BS) and its temporal resolution by the operator $\hat D_{t, t + \bar \tau}=\int \int dt_A dt_B e^{-\frac{(t_A-t)^2}{\Delta^2}}e^{-\frac{(t_B-t-\bar \tau)^2}{\Delta^2}}\ket{t_A, t_B}\bra{t_A,t_B}$, where $\Delta$ (supposed to be equal for both detectors) is the temporal resolution of the detectors, and detection events occur centered at times $t$ in arm $A$ and $t+\bar \tau$ in arm $B$. Notice that when $\Delta \rightarrow 0$ we have a perfectly time resolved detection and when $\Delta \rightarrow \infty$ (much larger than the biphoton's temporal envelope) we have the time unresolved case. The coincidence detection probability in this framework can be computed from $P_c(\bar \tau, \tau)=\int dt \bra{\psi'}\hat D_{t, t + \bar \tau} \ket{\psi'}$ that leads to 

\be\label{coincidence}
P_c (\bar \tau, \tau)=\int \int dt dt_1 dt_2 e^{-\frac{(t_1-t)^2}{\Delta^2}}e^{-\frac{(t_2-t-\bar \tau)^2}{\Delta^2}}| JTA(t_1-\tau, t_2)-JTA(t_2,t_1-\tau))|^2
\ee

We'll now consider that the $JTA$ is separable in variables $t_{\pm}=t_1 \pm t_2$, $JTA(t_1,t_2)=f_-(t_-)f_+(t_+)$, where $f_{\pm}$ are normalized function (this is frequently assumed in experiments, in particular, this is indeed a very good approximation in our work as well as in \cite{Kwiat, chen_hong-ou-mandel_2019}, for instance. Eq. \eqref{coincidence} becomes, up to a constant :
\be\label{later}
\tilde P_c(\bar \tau, \tau)=  \int \int  dt_- dt_+ e^{-\frac{(t_-+\bar \tau)^2}{2\Delta^2}}| f_-(t_--\tau)f_+(t_+-\tau)-f_-(-t_--\tau)f_+(t_+-\tau))|^2,
\ee
and we can integrate over $t_+$ to obtain
\begin{eqnarray} \label{time}
&&\tilde P_c(\bar \tau, \tau)=  \int dt_-  e^{-\frac{(t_-+\bar \tau)^2}{2\Delta^2}}| f_-(t_--\tau)-f_-(-t_--\tau)|^2= \nonumber \\
&&\int  dt_-  e^{-\frac{(t_-+\bar \tau)^2}{2\Delta^2}}(| f_-(t_--\tau)|^2+|f_-(-t_--\tau)|^2-2{\rm Re}[f^*_-(t_--\tau)f_-(-t_--\tau)]).
\end{eqnarray}
We can equally well write \eqref{time} in terms of frequency variables:
\begin{eqnarray} \label{TF}
&&\tilde P_c(\bar \tau, \tau)=  \int \int \int \int dt_- d\omega_-  d\omega'_- d\omega''_- e^{-2\Delta^2\omega_-^2}e^{i\omega_- (\bar \tau+t_-)}(e^{-i(\omega'_--\omega''_-)\tau} \tilde f^*_-(\omega'_-)\tilde f_-(\omega''_-) \times\nonumber \\
&&(e^{-i(\omega'_--\omega''_-)t_-}+e^{i(\omega'_--\omega''_-)t_-})-2{\rm Re}[e^{i\omega'_-(t_--\tau)}e^{-i(-t_--\tau)\omega''_-}\tilde f^*_-(\omega'_-)\tilde f_-(\omega''_-)])
\end{eqnarray}
Integrating over $t_-$ leads to:
\begin{eqnarray} \label{TF2}
&&\tilde P_c(\bar \tau, \tau)=  \int \int    d\omega_-  d\omega'_- d\omega''_- e^{-2\Delta^2\omega_-^2}e^{i\omega_- \bar \tau}(e^{-i(\omega'_--\omega''_-)\tau} \tilde f^*_-(\omega'_-)\tilde f_-(\omega''_-) \times \\
&&(\delta(\omega_--\omega'_-+\omega''_-)+\delta(\omega_-+\omega'_--\omega''_-))-2{\rm Re}[\delta(\omega_-+\omega'_-+\omega''_-)e^{-i\omega'_-\tau}e^{i\tau\omega''_-}\tilde f^*_-(\omega'_-)\tilde f_-(\omega''_-)]),\nonumber 
\end{eqnarray}
where $\tilde{f_-}$ is the Fourier transform of $f_-$. Eq. \eqref{TF2} can be integrated over $\omega_-$:
\begin{eqnarray} \label{TF3}
&&\tilde P_c(\bar \tau, \tau)=  \int \int \int    d\omega'_- d\omega''_- (e^{-i(\omega'_--\omega''_-)\tau} \tilde f^*_-(\omega'_-)\tilde f_-(\omega''_-) e^{-2\Delta^2(\omega'_--\omega''_-)^2}\times \\
&&(e^{i(\omega'_--\omega''_-) \bar \tau}+e^{-i(\omega'_--\omega''_-) \bar \tau})-2e^{-2\Delta^2(\omega'_-+\omega''_-)^2}{\rm Re}[e^{-i(\omega'_-+\omega''_-) \bar \tau}e^{-i\omega'_-\tau}e^{i\omega''_-\tau}\tilde f^*_-(\omega'_-)\tilde f_-(\omega''_-)]).\nonumber 
\end{eqnarray}
We can see from \eqref{TF3} that the time unresolved limit can be reached in two ways: one of them is by integrating \eqref{TF3} in $\bar \tau$ and the other by letting $\Delta \rightarrow \infty$, which transforms the Gaussian function into a delta. This seems indeed reasonable. In both situations we obtain
\be \label{integration}
\tilde P_c( \tau) \propto    1 -\int d \omega_-{\rm Re}[e^{-2i\omega_-\tau}\tilde f^*_-(\omega_-)\tilde f_-(-\omega_-)]. 
\ee

The expression for the Wigner function, denoted as $W(\mu=0, \tau)$, as presented in \cite{douce_direct_2013} and experimentally measured in \cite{PhysRevLett.115.193602, Kurzyna:22}, captures the essence of the Hong-Ou-Mandel (HOM) experiment. In alternative terms, the HOM experiment corresponds to a cut along the $\mu=0$ axis in the phase space of the Wigner function associated with the biphoton wavefunction $f_-$.

Remarkably, the operations of frequency or time resolution do not alter this fundamental aspect. Both operations can be interpreted as spectral filtering applied to the biphoton wavefunction, as demonstrated in \cite{fabre_producing_2020}. This perspective aligns with the findings in \cite{QCoherenceTomography}, where such spectral filtering is leveraged in the context of quantum-optical coherence tomography. Consequently, the core feature represented by the measurement of the $\mu=0$ axis in the phase space of the Wigner function remains robust and unaffected by variations in frequency or time resolution, which can be seen as state's post-selection.

Indeed, by letting $\Delta \rightarrow 0$ we can see that the coincidence detection is equivalent to the $\mu=0$ axis (with varying $\tau$) of the Wigner function of a Schrödinger cat in time (in analogy with the spectral resolution that filters Schrödinger cats in frequency): Eq. \eqref{time} becomes, in this limit:
\begin{eqnarray} \label{time2}
&&\tilde P_c(\bar \tau, \tau)=  | f_-(-\bar \tau-\tau)-f_-(\bar \tau-\tau)|^2= \nonumber \\
&&(| f_-(-\bar \tau-\tau)|^2+|f_-(\bar \tau-\tau)|^2-2{\rm Re}[f^*_-(-\bar \tau-\tau)f_-(\bar \tau-\tau)]),
\end{eqnarray}
If $f_-$ is a Gaussian function $f_-(t)=Ne^{-\frac{t^2}{\delta^2}}$, Eq. \eqref{time2} can be written as
\be\label{time3}
\tilde P_c(\bar \tau, \tau) \propto  e^{-2\frac{(\bar \tau+\tau)^2}{\delta^2}}+e^{-2\frac{(\bar \tau-\tau)^2}{\delta^2}}-2e^{-2\frac{\bar \tau^2+\tau^2}{\delta^2}}.
\ee
This means that we have post-selected a quantum superposition of two states centered at $\pm \bar \tau$ each. However, the envelope of the interference term, the third term in the r.h.s. of Eq. \eqref{time3}, has a visibility given by $e^{-2\bar \tau^2/\delta^2}$ if one considers the HOM experiment with varying $\tau$. Interestingly, this term persists even in the case of perfect time resolution, and the reason for that is that the delay $\bar \tau$ provides which-way information in the usual HOM interference. We can also notice that the same interpretation is possible by considering that $\bar \tau$ is the parameter to be estimated, or the interference controlling parameter, and this is expected since the two time delays play the same role.

An important point is that for unit visibility, we can see that, indeed, if $\bar \tau=0$, $P_c(0,\tau)=0 ~\forall ~\tau$ (this is true for any $f_-$ as can be seen from Eq. \eqref{time2}). 

Finally, in the case of non-perfect time-resolved measurement and finite $\Delta$, Eq. \eqref{time} becomes, in the case of a initial Gaussian temporal wavefunction $f_-$: 
\begin{equation} \label{time4}
\tilde P_c(\bar \tau, \tau)= \int  dt_-  e^{-\frac{(t_-+\bar \tau)^2}{2\Delta^2}}(e^{-2\frac{(t_--\tau)^2}{\delta^2}}+e^{-2\frac{(t_-+\tau)^2}{\delta^2}}-2e^{-2\frac{(t_-^2+\tau^2)}{\delta^2}})
\end{equation}

This situation of imperfect resolution corresponds to a noisy scenario, as shown in  \cite{PhysRevLett.130.200602}, and as can be seen from Eq. \eqref{time4}. The interference term (last term in the r.h.s. of Eq. \eqref{time4}) is given, in the general case, after the integration in $t_-$, by $e^{-\frac{2\bar \tau^2}{(4\Delta^2+\delta^2)}}e^{-\frac{2 \tau^2}{\delta^2}}$, which is a noisier situation than the one described in \eqref{time3}. 

Therefore, we can reach the following conclusions: a good time resolution reduces the visibility of the interference term in the HOM interferometer. It  corresponds to a post-selection of a Schrödinger cat-like state in time with non-perfect visibility, which can be associated to a non-pure state, if only coincidence events are considered, and anti-coincidence is inferred from the previous. This occurs even in the case of perfect temporal resolution. For this reason, it is clear that such a measurement strategy will never permit reaching the QCR bound even in the ideal situation of zero noise and perfect visibility and another measurement setting should be adopted. 

\section{The case of a Schr\"odinger cat}

We'll analyze in details the case of a Schr\"odinger cat (SC)-like state, an entangled state of two non-degenerate close to monochromatic photons. This state can be described as  $\ket{\phi}_c=\frac{1}{\sqrt{2}}\int d\omega_- (f(\omega_--\omega_o)+f(\omega_-+\omega_o))\ket{\frac{\omega_-+\omega_+}{2}, \frac{\omega_--\omega_+}{2}}$, where $f(x)$ is a function of width $\sigma$ and $\omega_o \gg \sigma$. We will chose for simplicity here $f(x)=(2/\sigma^2\pi)^{1/4}e^{-x^2/\sigma^2}$, a Gaussian function, but other functions would lead to the same type of results in the limit considered. We notice that we chose to express the state in terms of the collective variable $\omega_-=\omega_1-\omega_2$, since we consider to be in the regime discussed in the main text where $\omega_+=\omega_p$ is a constant.

We also have that $\abs{\int f(\omega)}^2  d\omega = 1$. The Wigner function $W(\mu, \tau)$ associated to state $\ket{\phi}_c$, experimentally studied in \cite{chen_hong-ou-mandel_2019, PhysRevApplied.19.054092, Kwiat}, is a superposition of two Gaussian centered at points $\mu = \pm \omega_o$ of the time-frequency phase space (TFPS), and interference fringes appear symmetrically around the axis $\mu=0$ for varying $\tau$ of TFPS. The HOM experiment is a direct measurement of the Wigner function at this axis \cite{douce_direct_2013}, and this explains the observed beating at a frequency proportional to the distance between the two possible states of the SC, modulated by a Gaussian envelope. This distance also corresponds to the spectral ``size" of the cat, and it is close to the inverse of the frequency of oscillation of the interference fringes. The HOM profile ( a cut at at $\mu=0$ in the 
$(\mu,\tau)$ TFPS) of the SC is given by 
\be \label{pccat}
P_c(\tau)=\frac{1}{2}(1-V\cos{(2\omega_o\tau)}e^{-\tau^2 \sigma^2})
\ee
for $0 \leq V \leq 1$, where $V$ is the visibility. We neglected the effects of the Gaussian peaks since $\omega_o \gg \sigma$. At this point, we can also notice that no matter what even function of width $\sigma$ satisfying this property would lead to the same results, as is the case of the superposition of square functions experimentally produced and discussed in the main text. As for an odd function, the only difference would be a sign change, such that $P_c(\tau)=\frac{1}{2}(1 +V\cos{(2\omega_o\tau)}e^{-\tau^2 \sigma^2})$.
From \eqref{pccat}, it is straightforward to compute the FI $F(\tau)$, which is given by 
\be \label{FIcat}
F(\tau)= \frac{4V^2 e^{-2\tau^2 \sigma^2}(\omega_o^2\sin{(2\omega_o \tau)}+2\tau \sigma^2\cos{(\omega_o \tau)})^2}{1-V^2\cos^2{(\omega_o \tau)}e^{-2\tau^2 \sigma^2}}, 
\ee

By supposing that $\Delta^2 \hat \omega_- = 4 \omega_o^2 + 2 \sigma^2 \gg \sigma^2$, we can see that \eqref{FIcat} leads to $F(\tau) \approx 4V^2 \Delta^2 \hat \omega_-  e^{-2\tau^2 \sigma^2}$, which displays a quadratic dependence upon the visibility  $V$, and is proportional to the QFI, as $\mathcal{F}=\Delta^2\omega_-^2$ in this case. 
We recall that for $V=1$, the limit $F=\mathcal{F}$ can be reached at $\tau=0$ in the HOM experiment. We remark that in the unphysical situation of $\sigma \rightarrow \infty$, $F(\tau)=4 V^2 \Delta^2\hat \omega_-$ is constant in time, and always presents the same scaling (quadratic).

\section{A discussion on resources}

This section develops the results obtained in \cite{PhysRevLett.131.030801} in the context of the present contribution. Notice that here the evolution operator acts in one photon only - one arm -, so  the Hamiltonian that generates the evolution is given by $\hat \omega_1$ and is given by $\hat H= \hbar \hat \omega_1 \tau$. The HOM experiment can be used as a metrological device to probe time with a given precision. This precision is related to the sensitivity of states to displacements in the TFPS in a given direction $\tau$ (the parameter to be estimated). Because of the HOM symmetry properties, it reveals a cut in TFPS associated to the function $f(\omega_-)$, {\it i.e.}, to the $\omega_-$ variable. Thus, since $\hat \omega_- = \hat \omega_1 - \hat \omega_2$, it is clear that a displacement of $\tau$ generated by $\omega_1$ is not the most efficient one: the associated QFI is given by ${\cal F}=4\Delta^2 \hat \omega_1= 4\Delta^2 \omega = \Delta^2\omega_-$, where we supposed that $\Delta^2 \hat \omega_1=\Delta^2 \hat \omega_2=\Delta^2 \omega$. Using, for instance, a displacement generated by the operator $\hat \omega_-$ would lead to a QFI of ${\cal F}=4\Delta^2 \hat \omega_-$. So, the two situations are different with respect to how they exploit the variance of the wave-function $f(\omega_-)$. Nevertheless, they both can be related to this variance and precision can be expressed in terms of it in some cases, as we'll discuss. 

If we consider, for instance, a separable state, we have that $\langle \hat \omega_1 \hat \omega_2 \rangle = 0$ (if we restrict ourselves to pure states, which is the case here), and $2\Delta^2 \hat \omega_1 = 2\Delta^2 \hat \omega = \Delta^2 \hat \omega_-$. We have considered here a separable state but the important property for us is the relation between the variances of $\omega_-$ and $\omega_1$, so it may well be that for some entangled states this same relation is observed, as discussed in \cite{PhysRevLett.131.030801}.  

If now we consider a case as the one studied in the main text, {\it i.e.}, where the states are close to maximally correlated in variable $\omega_-$ and $\omega_p=\omega_+$ is fixed, so $\Delta^2 \hat \omega_+=0$, we have that $4 \Delta^2 \hat \omega_1 = \Delta^2\hat \omega_-$. So, while for a separable state, ${\cal F}_s= 4\Delta^2 \hat \omega_1= 2  \Delta_s^2 \hat \omega_-$ for an entangled one (maximally correlated in variable $\omega_-$) ${\cal F}_e= 4\Delta^2 \hat \omega_1=  \Delta_e^2 \hat \omega_-$. This is simply a different way to express the results of \cite{PhysRevLett.131.030801}: in this reference, we have fixed the value of $\Delta^2 \omega$ as the classical available resource, as is the number of photons, or intensity, in the usual metrological picture where the spectral properties of states are disregarded. We have analyzed metrological properties from this perspective, unveiling the role of frequency in quantum precision limits. 
 
 A very important point is that the only requirement to establish these relations is the presence or absence of correlations of a certain type, relating the variances of $\omega_-$ and $\omega_1$. The specific form of the spectrum is not important, even though it may play a role on the required bandwidth to achieve a given value of $\Delta^2\omega$ and $\Delta^2\omega_-$ \cite{fabre_parameter_2021}. Thus, using a Gaussian spectrum or a Schrödinger cat like one is not really relevant as far as the QFI is concerned: states with different values of $\Delta^2\omega$ are states using different classical resources, and comparing their metrological power, even if interesting from an absolute perspective, is not the main goal of our work. As a matter of fact, a state with a given value of $\Delta^2\omega$ is equivalent to one with a given intensity. We know that increasing the intensity of a classical state may lead to a better precision than the one obtained with a quantum state. Nevertheless, in quantum metrology, the main subject of interest (at least the one related to the quantum properties of the system) is, in general, the scaling of precision with this resource. 

So, there are of course experiments using larger classical resources that reach better precision than the one reached here, as the ones with larger Schr\"odinger cat like states (\cite{Kwiat}, for instance). Nevertheless, these experiments follow the same scaling of the QFI with the overall resources as the one predicted in \cite{PhysRevLett.131.030801}, in one hand, and the same scaling of the ratio FI/QFI as the one predicted in the present manuscript for Schr\"odinger cat like states. As a matter of fact, we can check that in \cite{Kwiat} the authors obtain a visibility of $\approx 95\%$ and a ratio $QFI/FI \approx 88\%$, which are numbers entirely compatible with our model.

\section{Function with the worst and the best scaling: discussion}

We notice that for near perfect visibility ($V=1$), the solutions discussed in the main text corresponding to the best and the worst scaling of $\widetilde F_{V} = {\rm max} F(V,\tau)$ with the visibility $V$ - the sinusoidal and the constant functions, respectively - lead to a constant FI. As a matter of fact, a constant FI can be expressed in terms of the Wigner function $W(0,\tau)$ as 

\be\label{const}
\frac{(W'(0,\tau))^2}{(1-W^2(0,\tau))}=2a. 
\ee

We can gain some intuition on the scaling of the $\widetilde F_{V}$ with the help of the derivative:  
\be\label{deriva}
\frac{\partial F(V, \tau)}{\partial V} = 2  \frac{F(V, \tau)}{V}\times \frac{1}{(1-V^2W^2(0,\tau))} \leq \frac{V {\cal F}}{(1-V^2W^2(0,\tau))}. 
\ee
We start by discussing the case of the worst scaling of $\widetilde F_{V}$ with $V$ when $V$ approaches $1$ (recall that for $V=1$ and $\widetilde F_{1} = {\cal F}$ in the considered case). For this, we first analyse the point $\tau=0$ for different visibilities. Notice that, as discussed in the main text, at this point, the FI suffers a discontinuity dropping down from ${\cal F}$ to $0$ when $V$ changes from $1$ to $V \neq 1$. Thus, we must analyze \eqref{deriva} at different points $\tau \neq 0$, since $F(V<1,0)=0$. It is clear that the scaling gets worse as $W(0,\tau_M)$ is different from zero, and that the best scaling is such that $F(V, \tau_M)=V^2 {\cal F}$, which is possible for $W(0,\tau_M)=0$. 
V
If we want to intuitively understand the worst scaling, we can analyse the points $\tau \neq 0$.  If $F(1,\tau \neq 0) \neq 0$, the FI changes with $V$ from some finite value to another one, $F(V < 1,\tau_m) \neq 0$. If this is the case, we're sure that $\widetilde F_{V<1} > 0$. However, we know that this is not the worst possible scaling, since at point $\tau=0$ we have that the FI goes from the QFI to $F(V <1,0)=0$. Thus, in order to observe the worst possible scaling, we cannot have $W(0,\tau\neq 0)\neq W(0,0)$ for any $\tau$, otherwise we do not observe the worst possible scaling. This is of course an unphysical situation where we would also have $W'(0,\tau)=0$, so at the end, $\tilde F_{1}=0$ ! However, this situation helps to intuitively understand the problem and looking for a function with the closest behavior to the one described by with a finite value of $W'(0,\tau)$ and $W''(0,\tau)$. We are looking for a function producing the lowest possible value of $\widetilde F_{V}$, and this corresponds to minimizing $W''(0,\tau_M)/W(0,\tau_M)$. This is the requirement that leads to the physically possible version of the situation discussed above, and the function that has properties which are closest to the required ones is a Gaussian, since it minimizes $W''(0,\tau) ~\forall ~\tau$ while keeping ${\cal F} \neq 0$ at $V =1$. An analogous discussion is presented in \cite{PhysRevA.72.052332}.

Now, in order to intuitively understand the best scaling is obtained for $\tau_M$ we can see that $F(V<1,\tau) \leq V^2 F (1,\tau)~\forall ~\tau$. So, at every point $\tau$, the best scaling of the FI is quadratic in $V$. For $V =1$, $\widetilde F_1=F(1,0)={\cal F}$. We know also that $\widetilde F_{V}=V^2{\cal F}$ is the best possible scaling. However, if $F(1,\tau)\neq(<) F(1,0)$, it is clear that there cannot exist a $\tau_M$ such that $F(V<1,\tau_M) = V^2 {\cal F}=V^2 F(1,0)$, since this would mean that the scaling at point $\tau_M$ is better than quadratic: in other words, $\widetilde F_{V<1}= F(V<1,\tau_M)$ cannot be larger than $V^2 F(1,\tau_M) < V^2 {\cal F}$, so $\widetilde F_{V<1}=F(V<1,\tau_M) <  V^2 {\cal F}$ and we don't have the best scaling. The only way to circumvent this is having $F(1,\tau)$ constant in $\tau$, so that a quadratic scaling at every point leads to a quadratic scaling of $\widetilde F_{V}$.

Having a constant FI leads to equation \eqref{const}, that can be derived once, leading to 
\be\label{CFI2}
W'(0,\tau)(W''(0,\tau)+2aW(0,\tau))=0.
\ee
There are two solutions to this equation: $W'(0,\tau)=0$ (constant Wigner function), which corresponds to the case of worst scaling discussed above, and a harmonic oscillator-like equation that leads to the solution $W(0,\tau)=\cos{(\sqrt{2a}\tau)}$. Again, this is a non-physical situation that serves as a guideline, and the superposition of different types of functions that create the analogous to Schrödinger cat-like states (as two Gaussians or, as discussed in the main part of the text, two squared functions) can lead to a very close to oscillatory behavior optimizing the scaling, as we have seen from the experiments shown in the main text. We should mention that constant FI and the associated oscillatory probability distribution were discussed in another context in \cite{PhysRevE.97.042110}.

\section{Birefringence and phase space displacements}

In the studied broadband experimental set-up, the joint spectral amplitude (JSA), or the spectral function, is given by Eq. (2.18) of \cite{malte2019}:
\be\label{giorgio}
\phi_{PM}(\omega_+,\omega_-)={\rm sinc}{[\frac{L}{c}(\omega_+\Delta n_m(\omega_+)-\omega_-(\frac{\Delta n_b}{2}+\frac{\omega_-}{2}\frac{d n}{d\omega})]},
\ee
where $\omega_+=\omega_p$ (the pump's frequency), $\Delta n_m$ is the modal birefringence, $\Delta n_b$ comes from the signal and idler birefringence and $\frac{d n}{d\omega}$ is the chromatic dispersion. The argument of \eqref{giorgio}  has maxima at points $\omega_-^{\pm} = -(\Delta n_b /2 \pm \sqrt{\Delta^2n/4+2\omega_- \frac{d n}{d\omega}\omega_+\Delta n_m(\omega_+)})/(\frac{d n}{d\omega})$. Also, we can see that $\phi_{PM}$ is symmetric with respect to point  $\Omega =  -\Delta n_b /(2 \frac{d n}{d\omega})$. This sets the maximum of the Wigner function, or the point where a perfectly symmetric state is found. As a consequence, $W(\Omega,0)=1$, and having $V=1$ involves detecting this point. Nevertheless, since the HOM can only detect the Wigner function at axis $W(0,\tau)$ of the TFPS, we can never reach perfect visibility. 

\section{Some properties of the experimentally studied wave-functions}

We provide in this section the detailed properties of the wavefunctions that were used in the experiments studied in the main text (Table in Fig. \ref{fig1} and Wigner function in Fig. \ref{fig2}).

 \begin{figure}[h]
    \includegraphics[width=\columnwidth]{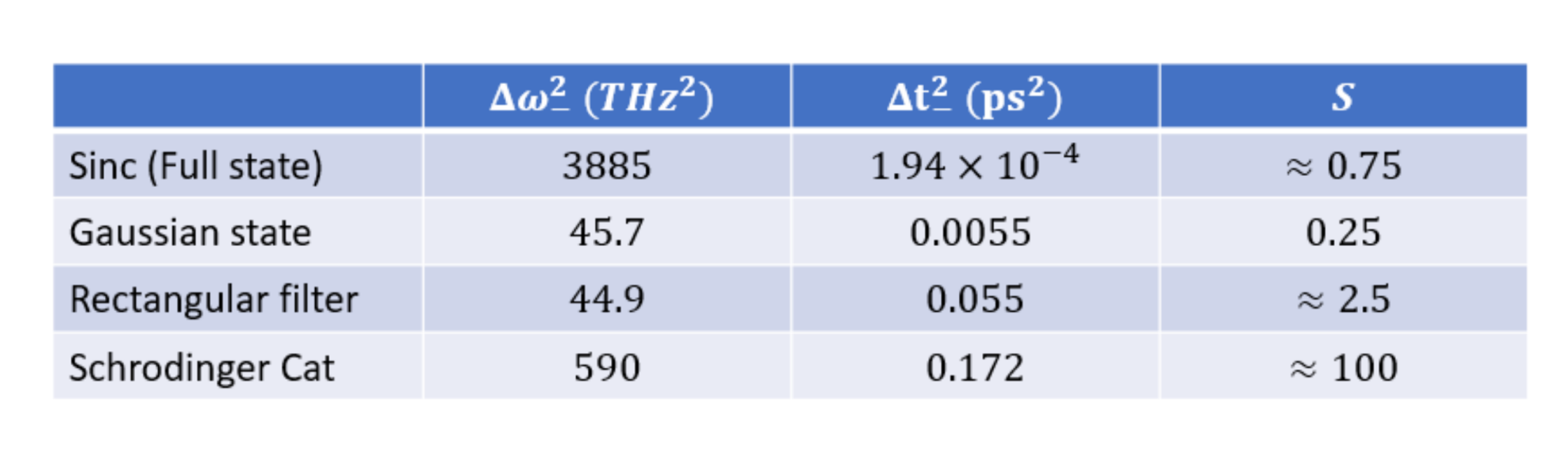}
    \caption{List of the functions experimentally produced and measured in the experiment presented in the main text, together with their phase space properties (the TFPS unit cell here is supposed to be {\cal S}=1/4. }
    \label{fig1}
\end{figure}

 \begin{figure}[h]
    \includegraphics[width=\columnwidth]{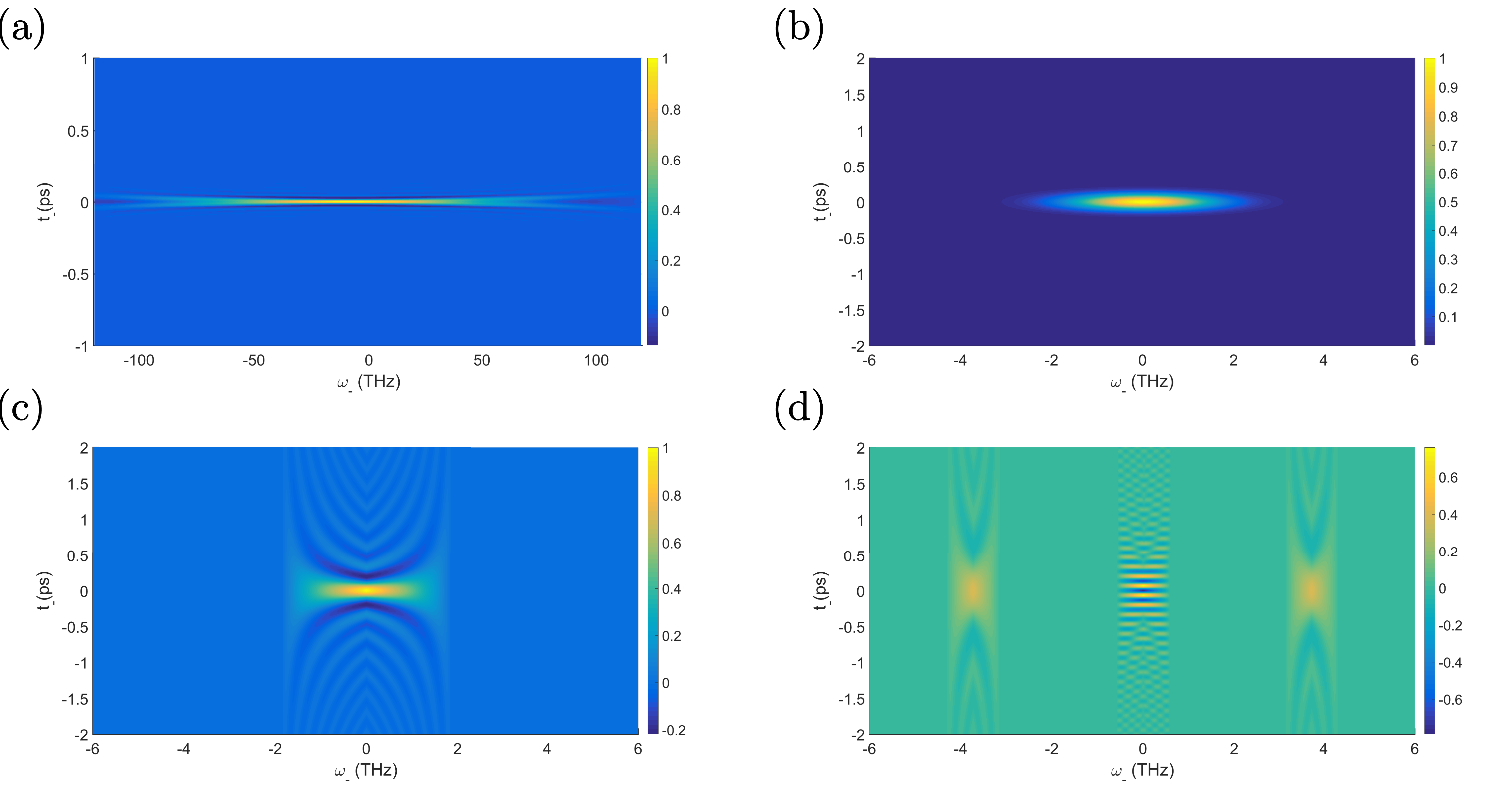}
    \caption{Wigner function of the four different biphoton states associated to variable $\omega_-$ : a) Sinc wavefunction b) Gaussian wavefunction c) Rectangular wavefunction d) Schrödinger cat-like wavefunction. The HOM experiment probes $W(\omega_-=0,\tau)$ ($\omega_-=0$ axis of the phase space) }
    \label{fig2}
\end{figure}

We stress that different spectral wave-function tailoring techniques exist  (see \cite{Taylor, boucher_toolbox_2015, Banaszek}, for instance), so as to further explore the different scaling properties of quantum states with visibility.

\section{Modeling visibility}

We'll now discuss the visibility model presented in the main text and experimentally implemented as well as other possible experimental and theoretical situations where this model applies. For such, we start by describing the situation where the two-photon wave-function is perfectly symmetric when the two photons are exchanged, which is a particular case of having two photons which are perfectly undistinguishable. The HOM experiment records interference between the situation where both photons are reflected and the one where both photons are transmitted. If the two photons are indistinguishable, as we consider here, then we can reach perfect visibility. Otherwise, if some external parameter creates distinguishability between the two possibilities, then the visibility is not perfect. 

In the experimental set-up discussed in the main text, we have a very close to indistinguishable state and distinguishability can be created in a controlled way by modifying the polarization of one of the photons with respect to the other. In order to understand this, we can write the general state that impinges into the HOM beam-splitter as
\be\label{input}
\ket{\psi}=\int \int d\omega_1d\omega_2 e^{i\omega_1 \tau}f(\omega_1,\omega_2)\ket{\omega_1,\omega_2},
\ee
where we considered that a time delay $\tau$ was applied at arm $1$ of the interferometer. After the beam-splitter, the part of the state that leads to coincidence detection is given by
\be\label{output}
\ket{\psi_c}=\frac{1}{\sqrt{2}} \int \int d\omega_1d\omega_2 \left (e^{i\omega_1\tau} f(\omega_1,\omega_2)\ket{\omega_1,\omega_2}\ket{T}-  e^{i\omega_2\tau}f(\omega_2,\omega_1)\ket{\omega_1,\omega_2}\ket{R}\right ),
\ee
where we have introduced the auxiliary states $\ket{T(R)}$ that can represent a which-way detector and model the visibility properties of the state. We can see that the coincidence detection probability $P_c(\tau)= \lVert \ket{\psi_c}\rVert^2$ is given by 
\be\label{coinci}
P_c(\tau)=\frac{1}{2}\left (1-{\rm Re}[\langle R|T\rangle   \int \int d\omega_1d\omega_2 \left (e^{i\omega_-\tau} f(\omega_1,\omega_2) f^*(\omega_2,\omega_1)\right )]\right ).
\ee
If $f(\omega_1,\omega_2)$ is a perfectly symmetric state with respect to the exchange of modes, we have that the visibility is given by ${\cal V}=\langle R|T\rangle$ (notice that $\ket{R}$ and $\ket{T}$ are not necessarily orthogonal to each other, it's precisely their overlap that determines wether the two paths of the interferometer are distinguishable or not). Notice that this is very close to the situation experimentally studied in the main text: the two produced photons have orthogonal polarizations, so state \eqref{input} can be written as
\be\label{input2}
\ket{\psi}=\int \int d\omega_1d\omega_2 e^{i\omega_1 \tau}f(\omega_1,\omega_2)\ket{\omega_1,\omega_2}\ket{H_1V_2},
\ee
where $\ket{H_i}$ and $\ket{V_i}$ are orthogonal polarizations associated to the $i$-th input spatial mode. We can use a wave-plate in arm, say $2$, so as to perform the transformation $\ket{V_2}\rightarrow \cos{\theta}\ket{V_2}+\sin{\theta}\ket{H_2}=\ket{\theta_2}$. By doing so, and using the beam-splitter mode transformation, we have that Eq. \eqref{output} can be written as 
\be\label{polar}
\ket{\psi_c}=\frac{1}{\sqrt{2}} \int \int d\omega_1d\omega_2 \left (e^{i\omega_1\tau} f(\omega_1,\omega_2)\ket{\omega_1,\omega_2}\ket{H \theta}-  e^{i\omega_2\tau}f(\omega_2,\omega_1)\ket{\omega_1,\omega_2}\ket{\theta H}\right ),
\ee
where we dropped the interferometer output spatial mode labels since they are redundant with the position in the kets. We have thus that the probability of coincidences is given by
\be\label{coco}
P_c(\tau)=\frac{1}{2}\left (1-{\rm Re}[\langle \theta H|H \theta\rangle   \int \int d\omega_1d\omega_2 e^{i\omega_-\tau} f(\omega_1,\omega_2) f^*(\omega_2,\omega_1)]\right ).
\ee
and the visibility is given by ${\cal V}=\langle \theta H|H \theta\rangle =\sin^2 \theta$.

The presented model relies on the fact that we have a very close to perfectly symmetric state, so that $ \int \int d\omega_1d\omega_2 f(\omega_1,\omega_2) f^*(\omega_2,\omega_1) \simeq 1$. If this is not the case, the visibility can still be controlled by changing the angle $\theta$ of the wave-plate, but it will be given by ${\cal V}= \langle \theta H|H \theta\rangle   \int \int d\omega_1d\omega_2  f(\omega_1,\omega_2) f^*(\omega_2,\omega_1)$ instead. In this case, for a general time $\tau$ we'll have
 \begin{eqnarray}\label{coco2}
&&P_c(\tau)=\frac{1}{2}\left (1-{\rm Re}[\langle \theta H|H \theta\rangle   \int \int d\omega_1d\omega_2 e^{i\omega_-\tau} f(\omega_1,\omega_2) f^*(\omega_2,\omega_1)]\right )=\\
&&\frac{1}{2}\left (1-\frac{{\cal V}}{ \int \int d\omega_1d\omega_2  f(\omega_1,\omega_2) f^*(\omega_2,\omega_1)}  {\rm Re}[ \int \int d\omega_1d\omega_2 e^{i\omega_-\tau} f(\omega_1,\omega_2) f^*(\omega_2,\omega_1)]\right).\nonumber
\end{eqnarray}
Notice that this approach has the advantage of not supposing any particular form for the wave-function $f(\omega_1,\omega_2)$ while allowing for studying the dependency of the FI with the visibility. 

In the main text we have modeled our results using this approach when we studied the wave-function naturally produced by the two-photon source. As for the other engineered functions, since by combining filtering techniques and state manipulation we have managed to reach very high values of visibility, we have used a different approach. By calling ${\cal V}_m$ the maximal visibility reached when both photon's polarization are the same ($\theta=\pi/2$), we have that 
\be\label{visel}
{\cal V}_m=\int \int d\omega_1d\omega_2  f(\omega_1,\omega_2) f^*(\omega_2,\omega_1), 
\ee
and we'll model $\int \int d\omega_1d\omega_2  f(\omega_1,\omega_2) f^*(\omega_2,\omega_1)e^{i\omega_-\tau} = {\cal V}_m \int \int |F_t(\omega_-)|^2e^{i\omega_-\tau} d\omega_1d\omega_2$,
where $F_t(\omega_-)$ is a target function, which is roughly the symmetric part of $f(\omega_1,\omega_2)$ with respect to the exchange of the two photon's spatial modes. Since the visibility reaches up to $99.5\%$, the error we make in the definition of the symmetry of the wave-function is lower that $0.25\%$, which corresponds to the probability of having an anti-symmetric wave-function. As explained in Section IV of this supplementary material, in the present device the parity defect comes from a displacement in the time-frequency phase space. As a matter of fact, we can show that for functions as Gaussians and Schrödinger cat-like states this displacement fits well the visibility model for the chosen wave-function parameters.

\bibliography{biblioMetro2}

\begin{thebibliography}{70}
\bibitem [{Com({\natexlab{a}})}]{CommentOdd}%
  \BibitemOpen
  \href@noop {} {\bibinfo {title} {We could have chosen it to be an odd
  function as well, and in the discussion that follows we would work with an
  anti-symmetric function rather than a symmetric one.}}
  ({\natexlab{a}})\BibitemShut {NoStop}%
\bibitem [{Com({\natexlab{b}})}]{Comment2}%
  \BibitemOpen
  \href@noop {} {\bibinfo {title} {Entangled non-degenerate photons are also
  useful to produce broadband states in the collective variable $\omega_-$
  using single photons with relatively narrow spectrum (before being entangled
  with one another).}} ({\natexlab{b}})\BibitemShut {NoStop}%
\end{thebibliography}

\appendix

\end{document}